\documentclass[aps,twocolumn,showpacs,preprintnumbers,amsmath,amssymb]{revtex4}

\usepackage{graphicx}
\usepackage{dcolumn}
\usepackage{bm}
\usepackage{color}

\begin{document}


\title{Superfluid to Mott insulator transition in the one-dimensional Bose-Hubbard model for arbitrary integer filling factors}
\author{Ippei Danshita$^{1}$}
\author{Anatoli Polkovnikov$^{2}$}
\affiliation{
{$^1$Computational Condensed Matter Physics Laboratory, RIKEN, Wako, Saitama 351-0198, Japan}
\\
{$^2$Department of Physics, Boston University, Boston, MA 02215, USA}
}

\date{\today}

\begin{abstract}
We study the quantum phase transition between the superfluid and the Mott insulator in the one-dimensional (1D) Bose-Hubbard model. Using the time-evolving block decimation method, we numerically calculate the tunneling splitting of two macroscopically distinct states with different winding numbers. From the scaling of the tunneling splitting with respect to the system size, we determine the critical point of the superfluid to Mott insulator transition for arbitrary integer filling factors. We find that the critical values versus the filling factor in 1D, 2D, and 3D are well approximated by a simple analytical function. We also discuss the condition for determining the transition point from a perspective of the instanton method.

\end{abstract}

\maketitle
\section{Introduction}
Systems of cold atoms in optical lattices have provided a highly controllable testing ground for quantum many-body physics~\cite{bloch-08}. Particularly, a transition from superfluid (SF) to Mott-insulator (MI) has attracted much attention. The SF to MI transition can be induced by increasing the depth of the optical lattice potential, and has been experimentally realized in 1D~\cite{stoeferle-04,fertig-05,mun-07,haller-10}, 2D~\cite{spielman-07,spielman-08,gemelke-09,bakr-10}, and 3D~\cite{mun-07,greiner-02,trotzky-10}.

It has been established that a system of cold bosonic atoms in an optical lattice can be quantitatively described by the Bose-Hubbard (BH) model~\cite{fisher-89,jaksch-98},
\begin{eqnarray}
\hat{H}=-J\sum_{\langle j,l \rangle}^L\left(\hat{b}_j^{\dagger} \hat{b}_{j+1} + h.c. \right)
+\frac{U}{2} \sum_{j}^L \hat{n}_j(\hat{n}_j -1),
\label{eq:BHH}
\end{eqnarray}
when the lattice is sufficiently deep, i.e. in the tight binding regime. Here $\hat{b}_j$ annihilates a boson at the lowest level localized on the $j$th site and $\hat{n}_j$ is the number operator.
The hopping energy $J$ corresponds to the overlap integral of two nearest-neighboring Wannier orbitals, and decreases exponentially when the lattice depth increases. The onsite interaction energy $U$ increases algebraically with the lattice depth. There is another important parameter, the number of particles per site $\nu$ (also called as filling factor) that is implicit in Eq.~(\ref{eq:BHH}). The ratio $U/(\nu J)$ controls the ground state phase of the BH model. When $U/(\nu J) \lesssim 1$, the SF phase is favored at any filling factors. When $U/(\nu J)$ increases at an integer filling factor such that $U/(\nu J) \gtrsim 1$, quantum fluctuations drive the system to the MI phase.

Since the SF to MI transition is one of the most remarkable phenomena emerging in the BH model, many previous studies have made efforts to determine the transition point both numerically~\cite{barbara-07,barbara-08,freericks-96,teichmann-09,kuhner-98,zakrzewski-07,ejima-11} and experimentally~\cite{mun-07,haller-10,spielman-08,trotzky-10}. Especially, theorists have accurately calculated the ratio $(U/J)_{\rm c}$ at the transition point for different filling factors and dimensionalities. In 2D and 3D, quantum Monte Carlo simulations have provided $(U/J)_{\rm c}$ at $\nu=1$~\cite{barbara-07,barbara-08} while the transition points at arbitrary filling factors have been calculated with use of the strong-coupling expansion (SCE) techniques~\cite{freericks-96,teichmann-09}. In 1D, $(U/J)_{\rm c}$ has been determined at low filling factors, namely at $\nu=1$~\cite{kuhner-98,zakrzewski-07,ejima-11} and $\nu=2$~\cite{zakrzewski-07, ejima-11}, using the quasi-exact numerical methods of density-matrix renormalization group (DMRG)~\cite{white-92} and time-evolving block decimation (TEBD)~\cite{vidal-04}. However, the transition points in the high filling region ($\nu\geq 3$) are yet to be obtained because of higher computational cost. Difficulty in treating the high filling region stems also from the fact that SCE fails to give an accurate transition point in 1D because the transition is of the Berezinski-Kosterlitz-Thouless (BKT) type.

In this paper, we calculate the critical points of the SF to MI transition in 1D for arbitrary integer filling factors, including the region of very high filling factors in which the BH model is equivalent to the $O(2)$ quantum rotor model~\cite{polkovnikov-05}. We emphasize that the determination of the transition points in the quantum rotor regime is important in the sense that the quantum rotor model effectively describes a regular array of Josephson junctions~\cite{sondhi-97} and liquid $^4{\rm He}$ absorbed in nanopores~\cite{yamamoto-08,yamashita-09,eggel-11}. In order to locate the transition point, one usually calculates static quantities such as the single-particle density matrix $\langle\hat{b}_j^{\dagger} \hat{b}_l\rangle$ and the density-density correlation function $\langle\hat{n}_j \hat{n}_l\rangle$~\cite{kuhner-98,zakrzewski-07,ejima-11}. In contrast, here we use a dynamic quantity, that is, energy splitting $\Delta$ caused by tunneling between two states with macroscopically distinct currents (or winding numbers)~\cite{kashurnikov-96}. In our previous work~\cite{danshita-10}, we have shown that the region of high filling factors can be efficiently treated with TEBD by imposing a lower bound for the occupation number at each site in addition to an upper bound. It has been also shown that the tunneling splitting can be accurately computed by means of TEBD. Using these prescriptions, we calculate $\Delta$ to determine the transition points. We find that the transition point as a function of $\nu$ is well approximated by 
\begin{eqnarray}
\left({U \over D \nu J}\right)_{\rm c} = a + b \nu^{-c},
\label{eq:critU}
\end{eqnarray}
where $D$ denotes the dimensionality of the system (e.g. $D=1$ for 1D), and the constants $a$, $b$, and $c$ are numerically determined. We show that one can express $(U/(D\nu J))_{\rm c}$ also in 2D and 3D, which has been obtained in Ref.~\onlinecite{teichmann-09}, in the form of Eq.~(\ref{eq:critU}) with different values of the constants. Given these results, one can immediately know the transition points for any filling factors and any realistic dimensions.

This paper is organized as follows. In Sec.~\ref{sec:model}, we introduce the system and the model considered here and explain how to determine the transition point from the tunneling splitting. We qualitatively discuss the relation between the tunneling splitting and the SF to MI transition from a perspective of the instanton method. In Sec.~\ref{sec:XXZ}, in order to examine the performance of the suggested procedure, we apply it to the 1D hardcore BH model with nearest-neighbor interactions that is exactly solvable by means of the Bethe ansatz. In Sec.~\ref{sec:UcVsNu}, we calculate the transition point in the BH model of Eq.~(\ref{eq:BHH}) as a function of the filling factor. 
In Sec.~\ref{sec:con}, we summarize the results.

\section{How to determine the transition point from the energy splitting}
\label{sec:model}
We consider a system of 1D lattice bosons in a ring-shaped geometry, i.e. with a periodic boundary. We assume for the moment that the system is in the SF phase. In the classical limit, which corresponds to the limit of $U/(\nu J) \rightarrow 0$ in the BH model, a state with a finite homogeneous current can be metastable, and its quasimomentum per particle is discretized as $p=2n\pi\hbar/(Ld)$, where the integer $n$ is the winding number, $L$ the number of lattice sites, and $d$ the lattice spacing. We suppose a situation that two states with different winding numbers, say $n_1$ and $n_2$, are degenerate. We define the winding number difference between the two states $n_{\rm d}$ as $n_{\rm d} \equiv |n_1 - n_2|$. When $n_{\rm d}\nu$ is an integer and $U/(\nu J)$ is finite, Umklapp scattering processes can induce phase slips via quantum tunneling to couple these two states. As a result, the degeneracy is broken and there emerges the energy splitting $\Delta$ that quantifies the tunneling rate. Applying the instanton techniques~\cite{coleman-85,sakita-85} to the phenomenological Tomonaga-Luttinger (TL) liquid model, Kashurnikov {\it et al.}~have derived the following scaling formula of the energy splitting at the smallest possible winding number difference $n_{\rm d}$ with respect to $L$~\cite{kashurnikov-96},
\begin{eqnarray}
\Delta \propto L^{-n_{\rm d}^2 K+1}
\label{eq:esp}
\end{eqnarray}
where the TL parameter $K$ is defined such that the effective action for the phase of the bosonic field $\theta(x,\tau)$ is described as
\begin{eqnarray}
S_{\rm TL} = \frac{\hbar K}{2\pi v}\int dx d\tau
\left[
(\partial_{\tau}\theta)^2
+v^2(\partial_{x}\theta)^2
\right].
\end{eqnarray}
Here $v$ is the sound speed.  

It is well-known that the SF phase is favored when $K>2/n_{\rm d}^2$ and that the SF to MI transition of the BKT type occurs at $K=2/n_{\rm d}^2$~\cite{giamarchi-04}. At the transition point, the renormalization of the TL parameter leads to a logarithmic correction on the scaling of $\Delta$ as~\cite{kashurnikov-96}
\begin{eqnarray}
\Delta \propto \frac{1}{L \ln L}.
\label{eq:espLog}
\end{eqnarray}
In the next section, we will see that the inclusion of the logarithmic correction is important to accurately calculate the transition point.

Given the scaling formulas of Eqs.~(\ref{eq:esp}) and (\ref{eq:espLog}), one can determine the transition point from the energy splitting as follows. First, using TEBD, we numerically compute $\Delta$ as a function of $L$ in the way described in Ref.~\onlinecite{danshita-10}. We next fit a function
\begin{eqnarray}
f(L)= {A L^{-B} \over \ln L}
\label{eq:fitF}
\end{eqnarray}
to the numerical data, where $A$ and $B$ are free parameters, and extract the exponent $B$. We determine the transition point from the condition that $B=1$.

Let us explain a reason why the exponent $B$ is equal to unity at the transition point from a different viewpoint, which is the relation between the SF to MI transition and the validity of the instanton formula of Eq.~(\ref{eq:esp}). An important point is that the instanton techniques used to derive Eq.~(\ref{eq:esp}) are based on the so-called dilute gas approximation (DGA), in which the instantons are assumed to be well separated from each other in the path-integral trajectories that contribute to the partition function~\cite{coleman-85,sakita-85}. In our system of 1D lattice bosons, since an instanton is regarded as a vortex in the space-time plane, the breakdown of DGA means that many vortices are created in the space-time coordinate so that they destroy the long-range coherence of bosonic phases, leading to the quantum phase transition to the Mott insulator. 
In short, the breakdown of DGA signals the Mott transition~\cite{danshita-11}. In general, the condition under which the dilute gas approximation is valid is that the size of an instanton along the imaginary-time axis $\tau_{\rm I}$ is much smaller than the tunneling time $\hbar/\Delta$. In the present case, since the winding number difference $n_d$ is of the order of unity, the instanton size is of the order of the system size, i.e. $\tau_{\rm I} \propto L$. Meanwhile, $\Delta$ is given by Eq.~(\ref{eq:esp}). Obviously, when $B>1$, the condition that $\tau_{\rm I} \ll \hbar/\Delta$ is held in the thermodynamics limit such that DGA is valid, i.e.~the system is in the SF phase. Thus, the condition for the Mott transition is given by $B=1$.

\section{Hardcore Bose-Hubbard model with the nearest-neighbor interactions}
\label{sec:XXZ}
In this section, we analyze the 1D hardcore BH model with the nearest-neighbor interactions,
\begin{eqnarray}
\hat{H}_{\rm hc}=-J\sum_{j=1}^L\left( e^{-i\theta} \hat{c}_j^{\dagger} \hat{c}_{j+1} + h.c. \right)
+ V \sum_{j=1}^L \hat{m}_j\hat{m}_{j+1},
\label{eq:hcBHH}
\end{eqnarray}
in order to illustrate that the critical point of the 1D superfluid to insulator transition can be accurately calculated along the procedure described in the previous section. Here, $V$ is the nearest-neighbor interaction, and $\hat{c}_{j}^{\dagger}$ and $\hat{m}_j$ are the creation and number operators of a hardcore boson at site $j$. We also include the phase twist $e^{-i\theta}$ in the hopping term in order to control the winding number of states. The hardcore constraint means that the maximum occupation number at each site is unity. This constraint leads to the identities between the operators of hardcore bosons and those of $1/2$-spins, namely $\hat{S}_j^z = \hat{m}_j -1/2$ and $\hat{S}_j^{-} = \hat{c}_j^{\dagger}$, which tell us that the model of Eq.~(\ref{eq:hcBHH}) is equivalent to the spin-$1/2$ XXZ model and is exactly solvable by means of the Bethe ansatz~\cite{takahashi-99}. According to the exact solution, there is a quantum phase transition between the SF and the density-wave insulator at half filling due to the competition between $V$ and $J$. When $V/J=0$, the particles favor to be delocalized and the system is in the SF phase. When $V/J$ increases, the Umklapp scattering tends to localize the particles, and the transition to the insulating state occurs at $V/J=2$. In the following, we show that our procedure provides a numerical value of the transition point that is very close to the exact one.

\begin{figure}[tb]
\includegraphics[scale=0.5]{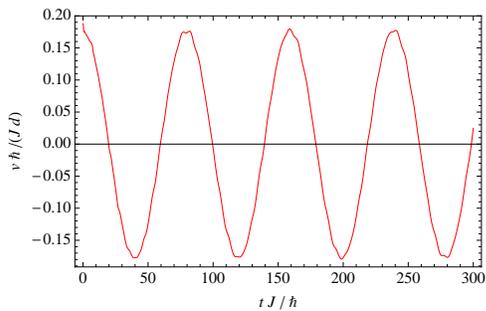}
\caption{\label{fig:vel}
(color online)  The time evolution of the current velocity $v(t)$ in the dynamics of the model of Eq.~(\ref{eq:hcBHH}), where $L=40$ and $V/J = 1.8$.
}
\end{figure}
\begin{figure}[tb]
\includegraphics[scale=0.5]{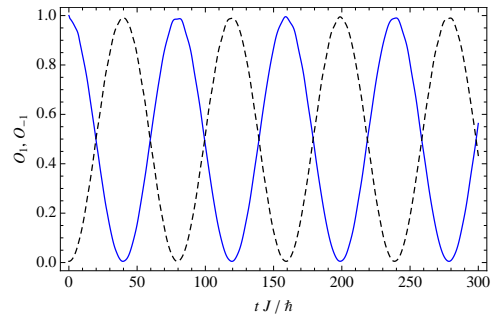}
\caption{\label{fig:ovlp}
(color online) The time evolution of the overlaps $O_1(t)$ (blue solid line) and $O_{-1}(t)$ (black dashed line) in the dynamics of the model of Eq.~(\ref{eq:hcBHH}), where $L=40$ and $V/J = 1.8$.
}
\end{figure}

Since the SF to MI transition occurs at $\nu=1/2$ in the model of Eq.~(\ref{eq:hcBHH}), the minimum winding number difference in possible phase slip processes is $n_{\rm d}=2$. In order to obtain the energy splitting for $n_{\rm d}=2$, we first prepare the ground state of Eq.~(\ref{eq:hcBHH}) with $\theta = 2\pi/L$, which is a state with winding number $n=1$. For dealing with our system with a ring-shape geometry, we use TEBD for a periodic boundary condition~\cite{danshita-09}. Taking the state with $n=1$ as the initial state and setting $\theta=0$, we next carry out the real-time evolution. Since two states with $n=1$ and $n=-1$ are degenerate when $\theta = 0$, it is expected that the supercurrent dynamics exhibit a coherent oscillation between these two states induced by quantum tunneling. To demonstrate this, we calculate the time evolution of the current velocity given by
\begin{eqnarray}
v=\frac{J d}{i\hbar N} \sum_j \langle \hat{c}_j^{\dagger}\hat{c}_{j+1} - h.c. \rangle,
\end{eqnarray}
where $N$ is the total number of particles. In Fig.~\ref{fig:vel}, $v(t)$ for $L=40$ and $V/J=1.8$ is shown. We clearly see that the velocity oscillates between $v(t=0)$ and $-v(t=0)$, i.e. between the states with $n=1$ and $n=-1$. We also calculate the overlap $O_n(t)=|\langle \Phi_n | \Psi(t)  \rangle|^2$ of the wave function with the ground state $|\Phi_n\rangle$ of the Hamiltonian (\ref{eq:hcBHH}) with $\theta = 2\pi n/L$. In Fig.~\ref{fig:ovlp}, we show overlaps $O_1(t)$ and $O_{-1}(t)$, which reconfirm that the wave function $|\Psi(t)\rangle$ coherently oscillates between $|\Phi_1\rangle$ and $|\Phi_{-1}\rangle$. We note that similar quantum-tunneling dynamics have been found also for quantum vortices in anisotropic traps~\cite{watanabe-07} and supercurrents in two-color optical lattices~\cite{nunnenkamp-08}.

Once the tunneling dynamics in real time are obtained, we can extract the energy splitting $\Delta$ by fitting $O_1(t)$  using the function
\begin{eqnarray}
g(t) = C \cos^2\left( \frac{\Delta}{2\hbar} t \right) + F
\end{eqnarray}
where $\Delta$, $C$, and $F$ are free parameters.

\begin{figure}[tb]
\includegraphics[scale=0.4]{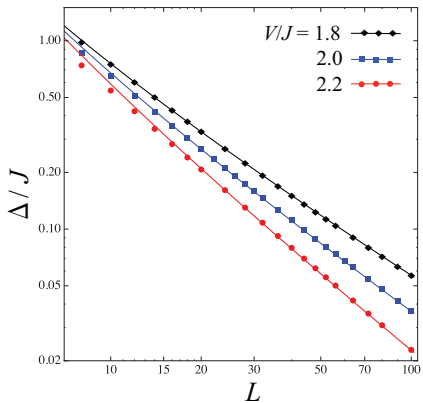}
\caption{\label{fig:delVsL}
(color online) The energy splitting versus the number of lattice sites $L$ for the 1D hardcore Bose-Hubbard model of Eq.~(\ref{eq:hcBHH}) at half filling. We take $V/J=1.8$ (black diamonds), $2.0$ (blue squares), and $2.2$ (red diamonds). The solid lines represent the best fits to the respective numerical date with the function of Eq.~(\ref{eq:fitF}), where $(A,B)=(2.80,1.11)$, $(2.28, 0.965)$, and $(1.84, 0.823)$ for $V/J=1.8$, $2.0$, and $2.2$. 
}
\end{figure}
\begin{figure}[tb]
\includegraphics[scale=0.45]{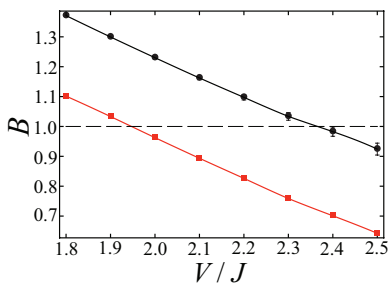}
\caption{\label{fig:beki}
(color online) The exponent $B$ versus $V/J$ for the 1D hardcore Bose-Hubbard model of Eq.~(\ref{eq:hcBHH}) at half filling. $B$ is extracted by fitting the numerical data of $\Delta$ versus $L$ upto $L=100$. We use the fitting function of Eq.~(\ref{eq:fitF}) with the logarithmic correction for red squares, while we use Eq.~(\ref{eq:fitFd}) without the correction for the black circles. The solid lines are guides to the eyes. The dashed line represents $B=1$, which is the condition to determine the transition point.
}
\end{figure}

In Fig.~\ref{fig:delVsL},  we plot the energy splitting between $|\Phi_1\rangle$ and $|\Phi_{-1}\rangle$ as a function of the number of lattice sites $L$ up to $L=100$. The numerical data are very well fitted by the function of Eq.~(\ref{eq:fitF}). From this fitting, we extract the exponent $B$ and calculate it varying $V/J$, as shown by the red squares in Fig.~\ref{fig:beki}. As expected, $B$ monotonically decreases with $V/J$. The condition that $B=1$ gives the transition point as $(V/J)_{\rm c} = 1.952 \pm 0.008$. This value is so close to the exact result, $(V/J)_{\rm c, exact} = 2$, that the relative error, $|(V/J)_{\rm c} - (V/J)_{\rm c, exact}| / (V/J)_{\rm c, exact}$, is smaller than $3\%$. 

We emphasize that our procedure provides an accurate value of the transition point even when the system size is relatively small. For instance, when we take the number of lattice sites up to $L=48$ for the fitting of the energy splitting, we find that $(V/J)_{\rm c} =1.925 \pm 0.013$ and that the relative error is still within $5\%$. This is a clear advantage of the use of the energy splitting over the correlation functions that require a substantially larger system size. We also note a disadvantage of our procedure that one needs a system with periodic boundaries in which the bipartite entanglement entropy is twice as large as that in a system with open boundaries.

In order to corroborate the necessity of the logarithmic correction in the fitting function of Eq.~(\ref{eq:fitF}), we instead use another function that does not include the correction,
\begin{eqnarray}
\bar{f}(L) = AL^{-B},
\label{eq:fitFd}
\end{eqnarray}
for the fitting. The black circles in Fig.~\ref{fig:beki} represent $B$ extracted with $\bar{f}(L)$, and the value of the transition point in this case is $(V/J)_{\rm c}=2.37\pm 0.02$. The relative error is $\sim 20\%$, which is much larger than the case with the logarithmic correction.

\section{The Bose-Hubbard model at arbitrary integer fillings}
\label{sec:UcVsNu}
\begin{table}[tb]
\begin{center}
\begin{tabular}{| c | c |}
\hline
$\nu$ & $(U/\nu J)_c$ \\
\hline
\,\, 1 \,\, & \,\, 3.128 $\pm$ 0.008 \,\, \\ 
\,\, 2 \,\, & \,\, 2.674 $\pm$ 0.010 \,\, \\
\,\, 3 \,\, & \,\, 2.510 $\pm$ 0.007 \,\, \\
\,\, 4 \,\, &  \,\, 2.420 $\pm$ 0.009 \,\, \\
\,\, 5 \,\, &  \,\, 2.374 $\pm$ 0.008 \,\, \\
\,\, 10 \,\, & \,\, 2.267 $\pm$ 0.008 \,\, \\
\,\, 20 \,\, & \,\, 2.213 $\pm$ 0.009 \,\, \\
\,\, 50 \,\, & \,\, 2.181 $\pm$ 0.008 \,\, \\
\,\, 100 \,\, & \,\, 2.172 $\pm$ 0.006 \,\, \\
\,\, 500 \,\, & \,\, 2.159 $\pm$ 0.005 \,\, \\
\,\, 1000 \,\, & \,\, 2.158 $\pm$  0.005 \,\, \\
\hline
\end{tabular}
\end{center}
\caption{\label{tab:1DUc}
The critical values $(U/(\nu J))_{\rm c}$ of the 1D BH model for several values of the filling factor $\nu$. 
}
\end{table}
\begin{figure}[tb]
\includegraphics[scale=0.5]{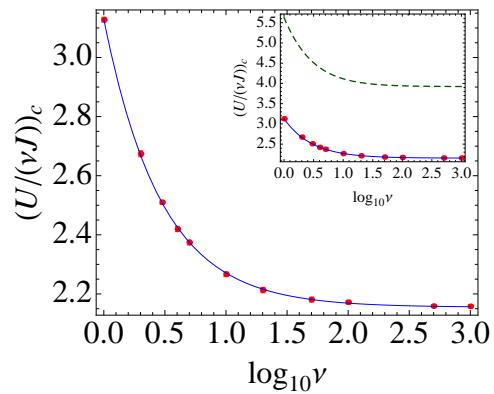}
\caption{\label{fig:1DUc}
(color online) The red circles represent the critical point $(U/(\nu J))_{\rm c}$ versus the filling factor. The solid line represents the best fit to the numerical data with the function of Eq.~(\ref{eq:critU}), where the numerical constants are obtained as $(a,b,c)=(2.16, 0.97, 2.13)$.  The transverse axis is depicted in a logarithmic scale. Notice that the critical point exists only at integer fillings although the solid line is continuous. In the inset, the same data are presented together with the transition point calculated with the analytical formula of Eq.~(\ref{eq:teich}) (green dashed line).
}
\end{figure}
In this section, we determine the critical point of the SF to MI transition of the 1D BH model (\ref{eq:BHH}) for arbitrary integer fillings, which is the main goal of the present paper. For this purpose, we start with the 1D BH model with a phase twist,
\begin{eqnarray}
\hat{H} \!=-\! J\sum_{j=1}^L\left( e^{-i\theta} \hat{b}_j^{\dagger} \hat{b}_{j+1} + h.c. \right)
+\frac{U}{2} \sum_{j=1}^L \hat{n}_j(\hat{n}_j -1).
\label{eq:BHH1D}
\end{eqnarray}
In the case of integer fillings, the minimum winding number difference in possible phase-slip processes is $n_{\rm d} =1$. In our previous work, we have shown a way to obtain the energy splitting for $n_{\rm d}=1$~\cite{danshita-10} using TEBD, which we follow in the present paper as well. We first calculate the ground state of Eq.~(\ref{eq:BHH1D}) with $\theta = 2\pi/L$ to obtain a state with $n=1$. Taking this state as an initial state and setting $\theta = \pi/L$, we calculate the real-time evolution. Since $|\Phi_1\rangle$ is  degenerate with $|\Phi_0\rangle$ at $\theta=\pi/L$, the dynamics exhibit a coherent oscillation between these two states. We extract the energy splitting from the oscillation, and calculate it as a function of the number of lattice sites up to $L=48$. The transition point is determined from the $L$-dependence of $\Delta$ as done in the previous section.

\begin{figure}[tb]
\includegraphics[scale=0.45]{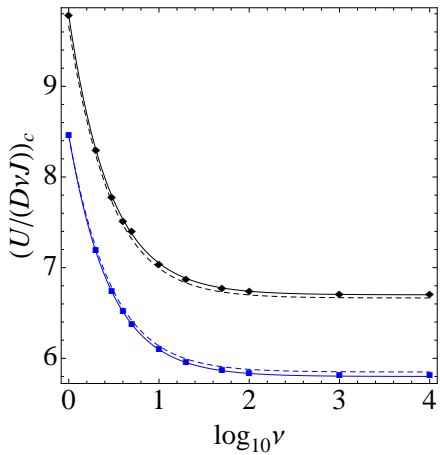}
\caption{\label{fig:2D3D}
(color online) The critical value $(U/(D \nu J))_{\rm c}$ versus the filling factor $\nu$ for 2D (blue squares) and 3D (black diamonds). The solid lines represent the best fit to the numerical data with the function of Eq.~(\ref{eq:critU}). The dashed lines represent the transition points calculated from the analytical formula of Eq.~(\ref{eq:teich}). The transverse axis is depicted in a logarithmic scale. The numerical data for 2D and 3D are taken from Ref.~\onlinecite{teichmann-09}. Notice that the critical point exists only at integer fillings although the solid line is continuous.
}
\end{figure}

In Table~\ref{tab:1DUc} and Fig.~\ref{fig:1DUc} (red circles), we show the critical point $(U/(\nu J))_{\rm c}$ as a function of $\nu$. Large scale DMRG analyses by Ejima {\it et al.} have provided the the most recent benchmark values of the critical points for $\nu = 1$ and $2$ as $(U/(\nu J))_{\rm c}=3.28$ and $2.78$ ~\cite{ejima-11}. Our results agree well with them to the extent that the relative deviation is within $5\%$. Since the ratio $U/(\nu J)$ quantifies the strength of quantum fluctuations in the quantum rotor limit ($\nu \rightarrow \infty$)~\cite{polkovnikov-05}, $(U/(\nu J))_{\rm c}$ is expected to be converged to a certain value when $\nu$ is sufficiently large. 
Indeed, the critical value monotonically decreases with $\nu$ and approaches an asymptotic value when $\nu \gg 1$. At $\nu = 1000$, $(U/(\nu J))_{\rm c}$ is well converged to the asymptotic value that corresponds to the quantum rotor limit. 

In Ref.~\onlinecite{teichmann-09}, it has been shown that an analytical formula accurately approximates the transition point for $D \ge 2$ as
\begin{eqnarray}
\left({D \nu J \over U}\right)_{\rm c} = \left({D \nu J \over U}\right)_{\rm c}^{\rm mf} 
+\frac{0.13\nu}{\sqrt{\nu(\nu+1)}D^{1.5}},
\label{eq:teich}
\end{eqnarray}
where
\begin{eqnarray}
\left({D \nu J \over U}\right)_{\rm c}^{\rm mf} = \nu^2+\frac{\nu}{2} - \nu \sqrt{\nu (\nu + 1)}
\end{eqnarray}
is the transition point obtained by a mean-field theory~\cite{fisher-89}. Although it has not been argued that this formula is valid for 1D, it is worth checking whether it works in 1D or not. The green dashed line in the inset of Fig.~\ref{fig:1DUc} represents the transition point given by Eq.~(\ref{eq:teich}). Obviously, the formula fails; the relative error is almost 100$\%$. This is not totally unexpected because the transition in 1D is special in the sense that it is of the BKT type while the transitions in higher dimensions are of the second-order. Instead of Eq.~(\ref{eq:teich}), we show that another analytical formula of Eq.~(\ref{eq:critU}) well approximates $(U/(\nu J))_{\rm c}$ versus $\nu$. As seen in Fig.~\ref{fig:1DUc}, the fitting with the function of Eq.~(\ref{eq:critU}) agrees with the data to the extent that the deviations are within the size of the data points.

\begin{table}[tb]
\begin{center}
\begin{tabular}{| c | c |}
\hline
Dimensionality: $D$ & $(a, b, c)$ \\
\hline
\,\, 1 \,\, & \,\, (2.16, 0.97, 2.13) \,\, \\ 
\,\, 2 \,\, & \,\, (5.80, 2.66, 2.19) \,\, \\
\,\, 3 \,\, & \,\, (6.70, 3.08, 2.18) \,\, \\
\hline
\end{tabular}
\end{center}
\caption{\label{tab:arbD}
The constants in the function of Eq.~(\ref{eq:critU}) obtained by the best fit to the numerical data for 1D, 2D, and 3D. 
}
\end{table}

The function of Eq.~(\ref{eq:critU}) is a good approximation also for the transition points in 2D and 3D. To show this, we depict in Fig.~\ref{fig:2D3D} the numerical data of $(U/(D \nu J))_{\rm c}$ versus $\nu$ for 2D (blue squares) and 3D (black diamonds), which are obtained using SCE in Ref.~\onlinecite{teichmann-09}, together with the best fit with the function of Eq.~(\ref{eq:critU}) represented by the solid lines. The transition points calculated from Eq.~(\ref{eq:teich}) are also plotted with the dashed lines. There we see that the function of Eq.~(\ref{eq:critU}) approximates the transition points as accurately as Eq.~(\ref{eq:teich}). The values of the numerical constants $(a, b, c)$ in the fitting function are summarized in Table~\ref{tab:arbD}.  

\section{Conclusions}
\label{sec:con}
In summary, we have studied the superfluid to Mott insulator transition in a system of one-dimensional lattice bosons.  We have shown that the transition points in 1D can be accurately determined from the energy splitting between two degenerate states with distinct winding numbers. We obtained the transition points for the Bose-Hubbard model with arbitrary integer filling factors, including the high filling limit corresponding to the quantum rotor regime. We have found a simple analytical formula of Eq.~(\ref{eq:critU}) that well approximates the transition point versus the filling factor for 1D, 2D, and 3D. With this formula, the transition points can be obtained easily and immediately for any fillings and any realistic dimensionalities.

\begin{acknowledgments}
The authors thank E. Altman, N. Prokof'ev, B. Svistunov, and S. Tsuchiya for stimulating discussions. I.~D. thanks Boston University visitors program for hospitality. A. P. was supported by NSF DMR-0907039, AFOSR FA9550-10-1-0110, and the Sloan Foundation. The computation in this work was partially done using the RIKEN Cluster of Clusters facility.
\end{acknowledgments}


\end{document}